\documentstyle[preprint,eqsecnum,aps]{revtex}
\newcommand{\be}{\begin{equation}}
\newcommand{\ee}{\end{equation}}
\newcommand{\bea}{\begin{eqnarray}}
\newcommand{\eea}{\end{eqnarray}}
\newcommand{\r}{\varphi}
\newcommand{\om}{\omega}
\newcommand{\p}{\partial}
\newcommand{\bs}{\!\!\!\!/}
\newcommand{\ran}[1]{\langle #1 \rangle}
\begin{document}
\draft
\title{\hbox{\hspace{20em} \normalsize  PUPT-1468, may 1994}  \Large Quantum
Field Theory Description of Tunneling in the Integer
Quantum Hall Effect}
\author{  D.\ G.\  Barci$^{a)}$ and L. Moriconi$^{b)}$}

\address{$^{a)}$ Departamento de F\'\i sica,
\\ Pontif\'\i cia Universidade Cat\'olica de Rio de Janeiro, \\
Caixa Postal 38071, 22452-970, Rio de Janeiro, Brasil.}
\address{$^{b)}$ Physics Department, Princeton University
\\Jadwin Hall, Princeton, NJ 08544, U.S.A.}
\maketitle
\begin{abstract}
We study the tunneling between two quantum Hall systems, along a quasi
one-dimensional interface. A detailed analysis relates microscopic parameters,
characterizing the potential barrier, with the effective field theory model for
the tunneling. It is shown that the phenomenon of fermion number
fractionalization is expected to occur, either localized in conveniently
modulated barriers or in the form of free excitations, once lattice effects are
taken into account. This opens the experimental possibility of an observation
of
fractional charges with internal structure, close to the magnetic length scale.
The coupling of the system to external gauge fields is performed, leading us to
the exact quantization of the Hall conductivity at the interface. The field
theory approach is well supported by a numerical diagonalization of the
microscopic Hamiltonian.
\end{abstract}

\newpage
\section{Introduction}
The relevance of the edges in the quantum Hall effect was stressed several
years
ago in a seminal paper by Halperin \cite{halperin}. The basic idea is that the
existence of gapless excitations at the edges, which follows from the general
principle of gauge invariance, provides a mechanism for the universal character
of the quantum Hall effect, or in other words, the presence of plateaus for the
Hall conductivity, independently of factors such as the degree of disorder and
the sample geometry. After this initial observation, the subject was left
untouched for a while until the works by Stone \cite{stone1} and
Wen \cite{wen1,wen2},
which showed many
interesting connections with the quantum field theory of (1+1)-dimensional
models
, in particular a relation to Kac-Moody algebras \cite{goddard}. One of the
consequences of these works is that they may be useful as a promising tool for
the study of tunneling between quantum Hall systems \cite {wen3}. Such a
research program is completely supported by the recent advances in the
fabrication of microstructures, which may give a real oppurtunity for testing
ideas originated from (1+1)-dimensional quantum field theory.

Motivated by this possibility, in this paper we will concentrate on the integer
quantum Hall effect in the simplest situation, viz, the tunneling along a quasi
one-dimensional interface, taking into account that both samples have their
first Landau levels completely filled. We will make contact here with the
theory
of fermion number fractionalization \cite {jackiw-rebbi,gw,ns}, which has found
some applications in the condensed matter physics of polymers \cite{s-s-h,j-s},
superfluid $^3He$ \cite{stone2}, and recently suggested to play an analogous
role at the interface of quantum Hall systems \cite{moriconi,barci-moriconi}.
In
fact, as we will show, similarly to the case of polyacetylene
\cite{s-s-h,j-s,takayama,s:1/3}, lattice effects may induce the presence of
fractional charges moving
along the interface. It is interesting, however, to observe that in the present
situation the excitations have a peculiar internal structure, composed by two
``subparticles", each one localized
around one of the edges.

This paper is organized as follows: in section II we present simple arguments
which
give the effective model of tunneling. This will be useful to show what is to
be expected from a more rigorous analysis, to be seen in further sections.
In section III
we build the effective model starting from the microscopic definition of an
interface. For the sake of a more complete analysis, and to study the
conductivity, we also consider the presence of additional external gauge
fields.
In section IV we explore some physical consequences of the model defined in the
previous section, such as charge trapping in specific barriers or via an
Aharonov-Bohm effect. The Hall conductivity is found to be quantized,
regardless
the presence of localized states or disorder at the interface. In section V we
study lattice effects in the tunneling, showing that lattice distortions may
occur,
associated to a gap
in the fermion spectrum as well as to fractionally charged excitations at the
interface. Distortions are represented by a complex bosonic field,
interacting with the fermion system. In section VI we perform a numerical
analysis of the problem of charge trapping in a modulated barrier,
obtaining a very good agreement between the microscopic system and the
effective (1+1)-dimensional field theory. In section VII we conclude our
discussion and point directions for future investigations.
Finally, in order to make the paper as self-contained as possible, there is an
appendix on fermion number fractionalization in field theory, where we show in
detail the computation of the fermionic current for the model of tunneling,
according to the method of Goldstone and Wilczek \cite{gw}.

\section{Heuristic Model}
We will show how the general form of the effective model of
tunneling may be found through phenomenological arguments, relying only on
a few basic assumptions \cite{moriconi}. This will provide us with some
motivation
before taking
the more complicated task of a complete microscopic analysis. In this section
we
assume the absence of two-body interactions and eletromagnetic
perturbations.

Our problem is to study what happens when we approximate two planar samples,
through a common plane,
taking into account that both
of them have their first Landau levels completely filled.
As one sample gets closer to the other, there will be some tunneling through
an approximately one-dimensional interface. A sufficient condition
for the presence of tunneling is that we have a certain degree of disorder
at the interface and that the distance between the samples be of the order
of the magnetic length $\ell$.

If we consider one of the samples as the square $-L/2 \leq x \leq L/2$
, $-L+\Lambda \leq y \leq \Lambda$, with $\Lambda \sim \ell$,
its edge can be defined by
$-\Lambda \leq y \leq \Lambda$.
The system is under the influence of a magnetic field $\vec B=B\hat z$
and its edge may be physically generated via the introduction of an electric
field $\vec E=E\hat y$,
which avoids the presence of electrons (at zero temperature)
in the region $y\geq \Lambda$. Working in the Landau gauge, $\vec A=(-By,0)$
, we have eigenfunctions localized only in the $y$ direction. According
to Stone \cite{stone1} we can define the charge density operator at the
edge as
\begin{equation}
j(x)=\int^\Lambda_{-\Lambda}\phi^+(x,y)\ \phi(x,y)\ dy \ , \label{em1}
\ee
where $\phi(x,y)$ is the field operator in second quantization, constructed
as a sum of only first Landau level states (which is a good approximation
to the case of strong magnetic fields). We expect the low energy
excitations at the edge to be associated to deformations of the quantum Hall
droplet, contained in $-\Lambda \leq y \leq \Lambda$ \cite{stone1,sakita}. The
Fourier expansion of
$j(x)$, as given by (\ref{em1}), may be recovered from
\begin{equation}
H=\int \psi^+_R(x)(-iv\p_x)\ \psi_R(x)\ dx \ , \label{em2}
\ee
a Hamiltonian of ``right-going'' chiral fermions. In this expression, $v=cE/B$
is
the drift velocity of electrons in the sample.
{}From (\ref{em2}) it follows that if the sign of $v$ is changed, the electric
current is inverted. In this case we would have a system of ``left-going''
chiral fermions, given by the field $\psi_L(x)$. This opposite situation
is exactly what occurs at the edge of the other sample. Therefore, the
transference of electrons between the samples will be given by the
introduction, in the Hamiltonian, of the tunneling operator $t\psi^+_R\psi_L$
(and its hermitian conjugate), where $t$ is the amplitude for the tunneling
of electrons across the interface. We claim, thus, that
\begin{equation}
H=\int\left[\psi^+_R(-iv\p_x)\ \psi_R+\psi^+_L(iv\p_x)\ \psi_L+
t\psi^+_R \psi_L+t^*\psi^+_L\psi_R\right] \ dx \ , \label{em3}
\ee
is the Hamiltonian which describes the tunneling process.
Taking $t\equiv-(t_1-it_2)$,
\begin{equation}
\psi=\left(
\begin{array}{l}  \psi_R \\
		  \psi_L
\end{array}         \right) \ , \ \label{em4}
\ee
and using the chiral representation for the $\gamma$ matrices,
\bea
\gamma^0&=&\sigma_1 \nonumber \\
\gamma^1&=&-i\sigma_2 \label{em5}\\
\gamma^5&=&\gamma^0\gamma^1=\sigma_3 \ , \ \nonumber
\eea
where $\sigma_1$, $\sigma_2$ and $\sigma_3$ are the Pauli matrices, it is a
simple matter to show that we can obtain (\ref{em3}) from the following
Lagrangian
\begin{equation}
{\cal L}=\bar\psi(i\gamma^0\partial_0+iv\gamma^1\partial_1)\psi +
\bar\psi(t_1+it_2\gamma^5)\psi  \ . \  \label{em6}
\ee
The fact that $t$ may depend on space and time variables, opens
interesting experimental possibilities,
related to the phenomenon of fermion number fractionalization. We will
come back to this point later.
\section{Microscopic Derivation}

In our phenomenological description, given by the Lagrangian (\ref{em6}),
the basic input is the tunneling amplitude $t$. A deeper question, therefore,
is to ask for a microscopic derivation of (\ref{em6}), in which the starting
point of analysis is the exact Hamiltonian of the system. Only in this way
we would know how to obtain the tunneling amplitude from a specific potential
barrier, characterizing the relevant class of
microscopic structures for the observation of interesting phenomena.

\subsection{COMPUTATION IN THE ABSENCE OF EXTERNAL GAUGE FIELDS}

Let us consider a system of independent electrons in the two-dimensional
$(x,y)$ plane, under the influence of a magnetic field $\vec{B}=B\hat z$ and
confined only in the $x$ direction by the strip $|x|<L/2$. We will be
interested in the limit $L\longrightarrow\infty$. A simple model of interface
is described by the Hamiltonian
\begin{equation}
H={1\over 2m}[(P_x+eBy)^2+P^2_y]+ eV(x,y) \ . \label{md1}
\ee
The first term in the Hamiltonian represents an
non-interacting two-dimensional electron gas.
The second term can be decomposed as the
sum of two contributions:
\begin{equation}
eV(x,y)=eV_1(y)+eV_2(x,y) \ , \
\ee
where
\bea
eV_1(y)&=&-Ay^2 \  ; \ A>0  \label{md2}  \\
eV_2(x,y)&=&g(x) \exp\left(-{y^2\over b^2}\right) \ . \ \label{md3}
\eea

The potential $V_1(y)$ is a parabolic barrier, which means, as we will see,
that for a certain range of the chemical potential $\mu$ there will be
a region $|y|<y_0$ completely free of electrons. By varying $\mu$ we can
have $y_0 \in (0, \infty)$. The other piece of $V(x)$, the term $V_2(x,y)$,
breaks translation invariance in the $x$ direction, generating a modulated
tunneling amplitude along the interface. It is clear that the Hamiltonian
(\ref{md1}) is unbounded from below, but this does not present any problem
in our approach: we could regularize the potential $V_1(y)$ to be
well-behaved for $|y| \gg \ell$, which would not change the physics of
tunneling
at the interface.

When $g(x)=0$, we have an exactly soluble model. In this case the
time independent Schr\"odinger equation is
\begin{equation}
\left\{{1\over 2m}\ [(P_x+eBy)^2+P^2_y]-Ay^2\right\}\ \r(x,y)=
E\r(x,y) \ .   \label{md4}
\ee
Imposing periodic boundary conditions in the $x$ direction, we can look for a
solution of the form $\r(x,y)=\exp(ik_nx)\ \xi(y)$, where
$k_n={2\pi n\over L}$, and $n$ is an integer. The equation for $\xi(y)$
is
\begin{equation}
\left\{{1\over 2m}\ [(k_n+eBy)^2+P^2_y]-Ay^2\right\}\ \xi(y)=
E\xi(y) \  ,    \label{md5}
\ee
that is
\bea
&&\left[{P^2_y\over 2m}+\left({(eB)^2\over 2m}-A\right)
\cdot \left(y+\left({(eB)^2\over 2m}-A\right)^{-1}\cdot {eB\over 2m} \
k_n\right)^2 +\right. \nonumber \\
&&\left.-{1\over 4}\ {(eB)^2\over m^2}\cdot \left({(eB)^2\over
2m}-A\right)^{-1} k^2_n+{1\over 2m}\ k^2_n\right]\ \xi(y)= E\xi(y)~.
\label{md6}
\eea
Therefore, equation (\ref{md6}) represents an one-dimensional harmonic
oscillator,
with eigenfunctions given by
\begin{equation}
\xi_{n,p}(y)=\exp\left[-{1\over 2\alpha ^2}\left(y-{\alpha ^4\over \ell^2}\
k_n\right)\right]\ H_p \left(y-{\alpha ^4\over \ell^2} \ k_n\right) \ ,
\label{md7}
\ee
where
\begin{equation}
\alpha =\left[2m\cdot\left({(eB)^2\over 2m}-A\right)\right]^{-{1\over 4}}=
\left[ 2m\cdot\left({1\over 2m\ell^4}-A\right)\right]^{-{1\over 4}} \label{md8}
\ee
and  $H_p$ is the Hermite polinomial of order $p$. The energies are
\begin{equation}
E_{n,p}=\left(p+{1\over 2}\right) \om_c+{k^2_n\over 2m} \cdot
\left[1-\left( {\alpha \over \ell}\right)^4\right] \ , \label{md9}
\ee
where  $\om_c={1\over m\ell^2}$ is the ciclotron frequency.

If the external magnetic field $B$ is high enough we can limit
the Hilbert space to the first Landau level. Therefore, from now on we will
restrict the state space to the set of normalized wavefunctions
\begin{equation}
\r_n(x,y)=\r_{n,0}(x,y)= \left( {1\over \pi\alpha^2\L^2} \right) ^{1/4}
exp\left\{ ik_n x-{1\over 2\alpha^2}(y-{\alpha^4\over \ell^2} k_n)^2 \right\}
\ . \ \label{md10}
\ee

The electric current associated with a function $\r_n(x,y)$ can be
immediately computed:
\bea
\langle J_y\rangle&=&{1\over m}\langle P_y\rangle =0  \label{md11} \\
\langle J_x\rangle&=&{1\over m}\langle P_x-{1\over \ell^2}\ y\rangle =
{k_n\over m}
\left[1-\left( {\alpha \over \ell}\right)^4\right] \  .\label{md12}
\eea

If we give a look at expression (\ref{md10}), we
easily recognize an interesting relation between the $k_n$ space and the
real $(x,y)$ space. The wavefunction $\r_n$ represents a delocalized state in
the $x$ direction and a localized state in the $y$ direction.
These states  are
centered in the $y$ direction around
$y_n=\frac{\alpha^4}{\ell^2}k_n$. From (\ref{md9}) and (\ref{md10}), one can
see that for
each value of the energy, there are two states, one centered at $+ y_n$
and the other at $- y_n$, so that the lesser the energy, the greater $|y_n|$.

We see from (\ref{md12}) that $n>0$ and $n<0$ define, respectively, electrons
moving to the right and to the left directions along the $x$ axis. We consider
$\alpha \stackrel{>}{_\sim} \ell$, which is equivalent of saying that the
wavefunctions are
spread in the $y$ direction whithin a region of order $\ell$. This means that
two wavefunctions, one with $n>0$ and the other with $n<0$ will have some
overlap only if their orbitals are distant from the line $y=0$ by a lenght of
order $\ell$. It is, thus, enough to consider $b=\ell$ in the expression for
$V_2(x,y)$, if one wants to produce a modulated tunneling amplitude at the
interface.

For each value of the chemical potential $\mu$ in the first Landau level,
there is a coordinate $y_0$ in such a way that all the states contained in the
region $|y|<y_0$ will be empty ones. The value of $y_0$ is a function of
$\mu$ and enters in our model as a phenomenological parameter.
Since $\alpha \stackrel{>}{_\sim} \ell$, it is adequate to
take $y_0 \sim \ell$ so that there is some
overlap between wavefunctions situated at the opposite sides of the
interface.

Acording to (\ref{md10}), the state centered at $y_0$ (or $-y_0$) defines a
value of $k_n$ (or $k_{-n}=-k_n$) given by $\bar{k}=y_0\ell^2/\alpha^4$. This
state has the quantum number $\bar{n}=\bar{k}L/2\pi$.
All we need to do from now on is to find a theory for the modes near
$\bar{k}$ and $-\bar{k}$.

Using the wavefunctions we found for the case of $g(x)=0$, given by
(\ref{md10}), we may
write the second quantized field operator as
\begin{equation}
\phi=\sum_{n<0} a^R_n\ \r_n(x,y)+\sum_{n>0} a^L_n\ \r_n(x,y) \ .\label{md13}
\ee
In this representation the Hamiltonian becomes
\begin{equation}
H_{eff}=\int\int \phi^+(x,y)\ H\phi(x,y)\ dydx \ , \ \label{md14}
\ee
where $H$ is the first quantized Hamiltonian (\ref{md1}).

In order to identify the filled states of our model with a ``Dirac sea'' in the
effective theory, it is necessary to make some redefinitions of the operators
$a_n^R$ and $a_n^L$. Let us make the following transformation
\bea
a_n^R&\longrightarrow a_{n+\bar{n}}^R \label{md15}\\
a_n^L&\longrightarrow a_{n-\bar{n}}^L \ . \ \label{md16}
\eea
The equation (\ref{md13}) now becomes
\begin{equation}
\phi=\sum_{n<\bar n} a^R_n\ \r_{n-\bar n}(x,y)+\sum_{n>-\bar n}
a^L_n\ \r_{ n+\bar n}(x,y) \ . \label{md17}
\ee
In the ``large box'' limit, the set of modes $k_n$ becomes dense and we
can define the continuum theory through
\bea
\sum_n&\longrightarrow& \frac{L}{2\pi}\int dk  \label{md18} \\
a_k\rightarrow\left(\frac{2\pi}{L}\right)^{1/2} a_k & \ , \
 & \r_k\equiv\r_n(L)^{1/2} \ .\ \label{md19}
\eea
We, thus, obtain
\begin{equation}
\phi=\left({1\over 2\pi}\right)^{1\over 2}\int_{k<\bar k}
a^R_k\ \r_{k-\bar k} \ dk +\left({1\over 2\pi}\right)^{1\over
2}\int_{k>-\bar k}  a^L_k\ \r_{k+\bar k} \ dk \ .\label{md20}
\ee

In order to completely identify the filled states with the
``Dirac sea'', it is necessary to redefine the energy too, setting to
zero the energy of the modes $\bar{k}$ and $-\bar{k}$. In this way, we shift
(\ref{md9}) to
\begin{equation}
E_k={(k^2-\bar k^2)\over 2m}
\left[1-\left( {\alpha \over \ell}\right)^4\right] -{(k^2-\bar k^2)\over
2\bar k}\ v  \ , \label{md21}
\ee
where
\begin{equation}
v=-{\bar k\over m} \left[1-\left( {\alpha \over \ell}\right)^4\right]>0\ .
\label{md22}
\ee

Substituting (\ref{md20}) into (\ref{md14}), we get
\bea
H_{eff}&=&\int_{k<\bar k} dk E_{k-\bar k}
a^{R^+}_ka^R_k+\int_{k>-\bar k} dk E_{k+\bar k}\ a^{L^+}_k\
a^L_k+       \nonumber \\
&+&{1\over 2\pi}\int_{k<\bar k \atop  k'>-\bar k} dkdk'\ dx\left\{
\exp\left[ i(-k+k'+2\bar k)x\right]\ g(x) c_1(k,k')\ a^{R^+}_ka^L_{k'} +
\hbox{H.c.} \right\}+ \nonumber \\
&+&{1\over 2\pi}\int_{k,k'<\bar k} dkdk'\ dx
\exp\left[ i(-k+k')x\right]\ g(x) c_2(k,k')\ a^{R^+}_ka^R_{k'} +\nonumber\\
&+&{1\over 2\pi}\int_{k,k'>-\bar k} dkdk'\ dx
\exp\left[ i(-k+k')x\right]\ g(x) c_2(-k,-k')\ a^{L^+}_ka^L_{k'} \ ,
\label{md23}
\eea
where
\begin{equation}
c_1(k,k')=\left({1\over \pi\alpha ^2}\right)^{1\over 2}\int dy
\exp\left\{-{1\over 2\alpha ^2}\left[\left(y-{\alpha ^4\over \ell^2}(k-\bar
k)\right)^2 + \left(y-{\alpha ^4\over \ell^2}(k'+\bar k)\right)^2\right] -
{y^2\over b^2}\right\} \label{md24}
\ee
and
\begin{equation}
c_2(k,k')=\left({1\over \pi\alpha ^2}\right)^{1\over 2}\int dy
\exp\left\{-{1\over 2\alpha ^2}\left[\left(y-{\alpha ^4\over \ell^2}(k-\bar
k)\right)^2 + \left(y-{\alpha ^4\over \ell^2}(k'-\bar k)\right)^2\right] -
{y^2\over b^2}\right\} \label{md25}
\ee
The above gaussian integrals can be computed exactly. We have
\begin{equation}
c_1(k,k')=\left({b^2\over \alpha ^2+b^2}\right)^{1\over 2}
\exp\left\{{\alpha ^6\over 2\ell^4}\left[{b^2\over 2(\alpha ^2+b^2)}(k+
k')^2 -((k-\bar k)^2+(k'+\bar k)^2)\right]\right\} \label{md26}
\end{equation}
\begin{equation}
c_2(k,k')=\left({b^2\over \alpha ^2+b^2}\right)^{1\over 2}
\exp\left\{{\alpha ^6\over 2\ell^4}\left[{b^2\over 2(\alpha ^2+b^2)}(k+
k'-2\bar k)^2- ((k-\bar k)^2+(k'-\bar k)^2)\right]\right\} \ . \  \label{md27}
\end{equation}
As we defined before, $b\simeq\alpha\simeq\ell$, so that we can write
\bea
c_1(k,k')&\simeq&\left({1\over 2}\right)^{1\over 2}
\exp\left\{{\ell^2\over 2}\left[{1\over 4}(k+
k')^2 -((k-\bar k)^2+(k'+\bar k)^2)\right]\right\} \label{md28}\\
c_2(k,k')&\simeq&\left({1\over 2}\right)^{1\over 2}
\exp\left\{{\ell^2\over 2}\left[{1\over 4}(k+
k'-2\bar k)^2 -((k-\bar k)^2+(k'-\bar k)^2)\right]\right\} \ . \  \label{md29}
\eea

The existence of tunneling depends essentially on the function $g(x)$. If,
for example $g(x)=\mbox{constant}$, the term in (\ref{md23}) representing
the tunneling between electrons $R$ and $L$ vanishes. This occurs because in
momentum space $\tilde g(k)=\delta(k)$, whereas in the tunneling process each
electron changes its momentum by approximately $2\bar k$ or $-2\bar k$.
Therefore, an interesting class of functions is given by $g(x)=e^{-2i\bar kx}
f(x) + c.c.$, where $f(x)$ is dominated in momentum space by modes
$\omega<<\bar k$. That is to say, writing $f(x)=\int exp(i\omega x)\tilde
f(\omega)\ d\omega$, we are considering functions centered at $\omega=0$,
with support in the interval  $\Delta\omega<<\bar k$. Substituting the proposed
form for
$g(x)$ into (\ref{md23}), we obtain
\bea
H_{eff}&=&\int_{k<\bar k} dk E_{k-\bar k}
a^{R^+}_ka^R_k+\int_{k>-\bar k} dk E_{k+\bar k}\ a^{L^+}_k\
a^L_k+ \label{md30}\\
&+& \int_{\om-\bar k<k<\bar k}dkd\om \big[\tilde f(\om)\ c_1(k,k-\om) \,
a^{R^+}_k a^L_{k-\om}+ \hbox{H.c.} \big]+\nonumber\\
&+&\int_{k<\bar k,k<-\bar k+\om} dkd\om \big[\tilde f(\om)\
c_2(k,k+2\bar k-\om)\ a^{R^+}_k a^R_{k+2\bar k-\om}+\hbox{H.c.}\big]+\nonumber
\\
&+&\int_{k>-\bar k,k>-3\bar k+\om} dkd\om \big[\tilde f(\om)\
c_2(-k,-k-2\bar k+\om)\ a^{L^+}_k a^L_{k+2\bar k-\om}+\hbox{H.c.}\big] \ , \
\nonumber
\eea
where, according to  (\ref{md28}) and (\ref{md29}),
\begin{equation}
c_1(k,k-\om)=\left({1\over 2}\right)^{1\over 2}
\exp\left\{-{\ell^2\over 2}\left[\left(k-{\om\over 2}\right)^2 +
2 \left(\bar k-{\om\over 2}\right)^2\right]\right\} \label{md31}
\ee
\begin{equation}
c_2(k,k+2\bar k-\om)=c_1(k,k-\om) \label{md32}
\ee
\begin{equation}
c_2(-k,-k-2\bar k+\om)=\left({1\over 2}\right)^{1\over 2}
\exp\left\{-{\ell^2\over 2}\left[\left(k-{\om\over 2}+2\bar k\right)^2 +
2 \left(\bar k-{\om\over 2}\right)^2\right]\right\} \ . \ \label{md33}
\ee

The expression (\ref{md31}) shows that for  $k \simeq \om \simeq 0$ we have
$c_1(k,k-\omega)\simeq\frac{1}{\sqrt{2}}e^{-(\bar k\ell)^2}$. The Hamiltonian
(\ref{md30}) describes the complete system, and we want to study only the modes
associated with tunneling, which are given approximately by $|k|<\bar
k$. We have to select in (\ref{md30}) only those degrees of freedom
involved in the tunneling process, retaining the most relevant couplings.
Therefore, we obtain from (\ref{md30}), a new effective Hamiltonian
\bea   \lefteqn{
H_{eff}^I=\int_{|k|<\bar k} dk E_{k-\bar k}
a^{R^+}_ka^R_k+\int_{|k|<\bar k} dk E_{k+\bar k}\ a^{L^+}_k\
a^L_k+\left({1\over 2}\right)^{1\over 2}\exp[-(\bar k\ell^2)] \times
}\nonumber \\
&\times&\int_{|k|<\bar k}dkd\om \big[\tilde f(\om)\ a^{R^+}_k
a^L_{k-\om} +\hbox{H.c.}\big] +c_2(-2\bar k,0)
\int_{|k|<\bar k} dkd\om \big[\tilde f(\om)\ a^{R^+}_{k-2\bar k+\om}
 a^R_k+\hbox{H.c.}\big]+\nonumber\\
&+&c_2(0,-2\bar k)\int_{|k|<\bar k} dkd\om \big[\tilde f(\om)\
 \ a^{L^+}_k a^L_{k+2\bar k+\om}+\hbox{H.c.}\big] +H' \  ,\label{md34}
\eea
where
\begin{equation}
H'=\int_{|k+2\bar k|<\bar k} dk E_{k-\bar k}
a^{R^+}_ka^R_k+\int_{|k-2\bar k|<\bar k} dk E_{k+\bar k}\
a^{L^+}_k\ a^L_k \  . \label{md35}
\ee
We can retain in $H^I_{eff}$, only the first three terms. In fact, using
(\ref{md32})  and (\ref{md33}) we have
\begin{equation}
c_2(-2\bar k,0)=c_2(0,-2\bar k)\simeq\frac{1}{\sqrt{2}}e^{-3(\bar k\ell)^2}
\ . \ \label{md36}
\ee
That is, the ratio between $c_2(-2\bar k,0)$  and the factor
$\frac{1}{\sqrt{2}}e^{-(\bar k\ell)^2}$  in the tunneling amplitude is
$e^{-2(\bar k\ell)^2}$. Now, if $\bar k\sim 1/\ell$,
we have
$e^{-2(\bar k\ell)^2} \sim e^{-2} \sim 10^{-1}$, which indeed shows that we can
keep only the first three terms of (\ref{md34}), to investigate the tunneling.

The expressions for $E_{k-\bar k}$ and $E_{k+\bar k}$, can be linearized
around $k=0$. Using (\ref{md21}) we get
\bea
E_{k-\bar k}&=& k v     \nonumber \\
E_{k+\bar k}&=& -k v \ . \    \label{md37}
\eea
Taking into account these approximations, we obtain the effective Hamiltonian
\bea
H_{eff}^I&=&\int_{|k|<\bar k} dk \ kv \
a^{R^+}_ka^R_k-\int_{|k|<\bar k} dk \ kv \ a^{L^+}_k\
a^L_k+\nonumber \\
&+&c\int_{|k|<\bar k}dkd\om \big[\tilde f(\om)\
 \ a^{R^+}_k a^L_{k-\om}+\hbox{H.c.}\big]\ , \label{md39}
\eea
where
\begin{equation}
c=\left({1\over 2}\right)^{1\over 2}\exp[-(\bar k\ell)^2]
=\left({1\over 2}\right)^{1\over 2}\exp\left[-\left({y_0\over
\ell}\right)^2 \right]\ . \label{md40}
\ee
Defining now
\bea
\psi_R(x)&=&{1\over \sqrt{2\pi}}\int dk \exp(ikx)\ a^R_k \label{md41}\\
\psi_L(x)&=&{1\over \sqrt{2\pi}}\int dk \exp(ikx)\ a^L_k \ , \label{md42}
\eea
we see that  $H^I_{eff}$ is, in coordinate space,
\begin{equation}
H^I_{eff}=\int dx \left[\psi^+_R(-iv{\p\over \p
x})\psi_R+ \psi^+_L(iv{\p \over \p x}) \psi_L+cf(x)\
\psi^+_R \psi_L
+cf^*(x)\ \psi_L^+\psi_R \right] \ , \  \label{md43}
\ee
which agrees with (\ref{em3}).

\subsection{INTRODUCTION OF GAUGE FIELDS}

We will obtain now the effective Hamiltonian for the
interface, taking into account the presence of an external gauge field
$a_\mu$. We have to consider the more general microscopic Hamiltonian
\begin{equation}
H={1\over 2m}\left[(P_x+eB_y-ea_1)^2+ (P_y-ea_2)^2\right] +eV(x,y)+
ea_0 \ . \label{gf1}
\ee
We can write
\begin{equation}
H=U^+H_0U \ , \ \label{gf2}
\ee
where
\begin{equation}
H_0={1\over 2m}\left[(P_x+eB_y)^2+ P_y^2\right] +eV(x,y)+
ea_0 \ ,  \label{gf2'}
\ee
and $U$ is the unitary operator
\begin{equation}
U=\exp\left(-ie\int^{\vec x}_{0,c} \vec a\cdot d\vec x\,'\right) \ .\label{gf3}
\ee
In this expression $c$ represents a path in the $(x,y)$ plane, given by
\begin{equation}
c:\left\{
\begin{array}{lc}
		y=0	\ , \	  0<x'<x   \\
		x=0     \ , \     0<y'<y
\end{array}
\right.			\label{gf4}
\ee
Let us supose that $a_\mu$ is a small static field with slow variations
in the magnetic length scale. From $H_0$, equation (\ref{gf2'}), we are
led, according to our previous computations, to the effective Hamiltonian
\bea
H^I_{0,eff}&=&\int dx\left[\psi^+_R\left(-iv{\p\over \p
x}\right)\psi_R+ \psi^+_L\left(iv{\p \over \p x}\right)\right.\psi_L+cf(x)\
\psi^+_R \psi_L
+cf^*(x)\ \psi_1^+\psi_R\bigg] \nonumber \\
&+&\int dx \ ea_0(x,y=0) \left[\psi^+_R\psi_R+\psi^+_L \psi_L\right] \ . \
\label{gf5}
\eea
The second quantized unitary operator $U$ takes the form
\begin{equation}
U\simeq 1-ie\int \phi^+(x,y) \left[\int^x_0 a_1(x',0)\ dx'+ \int^y_0
a_2(x,y') \ dy'\right] \ \phi(x,y)\ dxdy \ . \label{gf6}
\ee
Using (\ref{md20}) and neglecting terms similar to $a_k^{R^+}a_{k'}^L$, we
obtain
\bea
U&=&1-{ie\over 2\pi} \int _{k,k'<\bar k}dkdk'dx
\exp[i(-k+k')\ x]\int^x_0 a_1(x',0)\ dx'\ \tilde c_2(k,k') [a^{R^+}_k
a^R_{k'}+\nonumber\\
&+&a^{L^+}_{-k} a^L_{-k'}] -{ie\over 2\pi} \int_{k,k'<\bar k} dkdk'dx
\exp[i(-k+ k')\ x] \ c_3(x,k,k')\ a^{R^+}_k a^R_{k'}+\nonumber \\
&-&{ie\over 2\pi} \int_{k,k'<-\bar k} dkdk'dx \exp[i(-k+k')\ x] \
c_4(x,k,k') \ a^{L^+}_ka^L_{k'} \ , \  \label{gf7}
\eea
where
\bea
c_3(x,k,k')&\equiv& \left({1\over {\pi \alpha^2}}\right)^{1\over 2}
\int dy \int^y_0 a_2(x,y')\ dy' \exp\left\{-{1\over{ 2 \alpha^2}} \left[\left(
y-{ \alpha^4 \over \ell^2}(k-\bar k)\right)^2+\right.\right.\nonumber \\
&+&\left.\left. \left(y-{\alpha^4 \over \ell^2}(k'-\bar
k)\right)^2\right]\right\} \ , \  \label{gf8}
\eea
\bea
c_4(x,k,k')&\equiv& \left({1\over {\pi \alpha^2}}\right)^{1\over 2}
\int dy \int^y_0 a_2(x,y')\ dy' \exp\left\{-{1\over{ 2 \alpha^2}} \left[\left(
y-{ \alpha^4 \over \ell^2}(k+\bar k)\right)^2+\right.\right.\nonumber\\
&+&\left.\left. \left(y-{\alpha^4 \over \ell^2}(k'+\bar
k)\right)^2\right]\right
\} \label{gf9}
\eea
and $\tilde c_2(k,k') \equiv \lim_{b\rightarrow\infty}c_2(k,k')$, with
$c_2(k,k')$ given by (\ref{md27}). We must retain in equation (\ref{gf7}) only
the modes associated with
tunneling, $|k|<\bar k$. Approximating $a_2(x,y)$ by $a_2(x,0)$ and
substituting
$k=k'=0$ in the expressions for $\tilde c_2$, $c_3$ and $c_4$, we get
\bea
U^I&=&1-{ie\over 2\pi}\int_{|k|,|k'|<\bar k}dkdk' dx \exp
[i(-k+k')x] \int^x_0 a_1(x',0)\ dx' [a^{R^+}_k a^R_{k'}+\nonumber \\
&+&a^{L^+}_k a^L_{k'}]-{ie\over 2\pi}\int_{|k|,|k'|<\bar k}dkdk' dx \exp
[i(-k+k')x]\cdot (-y_0 a^{R^+}_k a^R_{k'}+\nonumber \\
&+&y_0a^{L^+}_k a^L_{k'}) \ a_2(x,0) \ , \label{gf10}
\eea
or alternatively, using the definitions (\ref{md41}-\ref{md42}),
\bea
U^I&=&1-ie\int
dx\left[\psi^+_R\psi_R\left(-y_0a_2(x,0)+\int^x_0 a_1(x',0)\ dx'\right)
+\right.\nonumber\\
&+&\left.\psi^+_L\psi_L\left(y_0a_2(x,0)+\int^x_0 a_1(x',0)\
dx'\right)\right]  \ . \ \label{gf11}
\eea
To compute $H^I_{eff}=U^{I^+}\ H^I_{0,eff} U^I$, it is important to know the
following operator products:

\bea   \lefteqn{
\hbox{{\it i})}\ U^{I^+} \psi^+_R(x)\ U^I=}\nonumber \\
&&\psi^+_R(x) +i e
\int dx' \left[\psi^+_R(x') \psi_R(x')\left(-y_0 a_2(x',0)+\int^{x'}_0
a_1(x{'}',0) dx{'}'\right)\psi^+_R(x)\right]\nonumber \\
&=&\psi^+_R+ie\left(-y_0a_2(x,0)+\int^x_0 a_1(x',0)\ dx'\right)\
\psi^+_R(x)\nonumber \\
&\simeq&\exp\left[ie\left(-y_0a_2(x,0)+\int^x_0 a_1(x',0)\ dx'\right)\right]\
\psi^+_R(x) \  \label{gf12}
\eea
and, in the same way,
\begin{equation}
\hbox{{\it ii})}\ U^{I^+}\ \psi^+_L(x)\
U^I=\exp\left[ie\left(y_0a_2(x,0)+\int^x_0 a_1(x',0)\
dx'\right)\right]\ \psi^+_L(x) \ . \label{gf13}
\ee
Therefore, we get, using the above relations,
\bea   \lefteqn{
\hbox{{\it iii})}\ U^{I^+}\left[\psi^+_R{\p\over \p x}\ \psi_R
 -\psi^+_L{\p\over \p x}\ \psi_L\right]\ U^I=}\nonumber\\
&&=\psi^+_R\left[{\p\over \p x}-ie\ {\p\over \p x}
\left(-y_0a_2(x,0)+\int^x_0 a_1(x',0)\ dx'\right)\right]\ \psi_R+\nonumber\\
&-&\psi^+_L\left[{\p\over \p x}-ie\ {\p\over \p x}
\left(y_0a_2(x,0)+\int^x_0 a_1(x',0)\ dx'\right)\right]\ \psi_L=\nonumber\\
&=&\psi^+_RD_x\psi_R-\psi^+_LD_x\psi_L+ie{\p a_2\over \p x}\ y_0
[\psi^+_R\psi_R +\psi^+_l\psi_L] \ , \label{gf14}
\eea
where
$D_x\equiv {\p\over \p x}-ie a_1(x,0) $, and also
\begin{equation}
\hbox{{\it iv})}\ U^{I^+}\psi^+_R\psi_LU^I=\exp(-2iea_2(x,0)\ y_0)\
\psi^+_R\psi_L \ .  \label{gf15}
\ee
{}From (\ref{gf14}) and (\ref{gf15}) we obtain
\bea
H^I_{eff}&=& U^{I^+}H^I_{0,eff}U^I=\int dx[-i\bar \psi \gamma ^1
vD_1\psi - c\bar \psi(f_1+if_2\gamma ^5) \exp(2iea_2y_0\gamma ^5) \psi+
\nonumber\\
&+&ev\ {\p a_2\over \p x}\ y_0\bar\psi\gamma ^1\gamma ^5\psi] \ . \label{gf16}
\eea
The gauge invariant Lagrangian associated with this
Hamiltonian is
\begin{equation}
{\cal L}=i\bar\psi(\gamma ^0D_0+v\gamma ^1D_1)\ \psi+ c\bar\psi (f_1+if_2
\gamma ^5)
\exp( 2iea_2y_0\gamma ^5)\psi+eva_2y_0\p_\mu j^\mu_5 \ . \label{gf17}
\ee

It is not dificult to understand why the above Lagrangian is the correct
generalization of (\ref {em6}), which takes into account the presence of
gauge fields. The first attempt, in order to implement gauge invariance,
would be to replace, in (\ref {em6}), $\gamma^\mu \partial_\mu$ (we are now
considering
$v=1$) by the covariant derivative, $D\bs= \gamma^0 ( \partial_0 +
iea_0) + \gamma^1 (\partial_1 + iea_1)$. This is, however, only a partial
answer. It is necessary to consider that the operators $\psi_R(x)$
and $\psi_L(x)$ create hole states in different positions of the
two-dimensional
plane. In this way, if a gauge transformation is performed, $\psi_R$ and
$\psi_L$ will be multiplied by different phase factors. Besides that, one
additional
requirement is that the effective model of tunneling be gauge invariant at
a classical level. This condition is obtained from the decoupling betweeen the
modes at the interface and the bulk degrees of freedom. Considering
the gauge transformation (below $\mu = 0,1$)
\begin{equation}
a_\mu \rightarrow a_\mu+\partial_\mu\alpha~~~ \mbox{and}~~~
a_2\rightarrow a_2+\partial_2\alpha|_{y=0}\ , \label{gf18}
\end{equation}
associated to
\begin{equation}
\psi\rightarrow
exp\left[i\left(\frac{\phi}{2}\gamma^5-e\alpha\right)\right]\psi
{}~~~\mbox{and}~~~
\bar\psi\rightarrow \bar\psi exp\left[i\left(\frac{\phi}{2}\gamma^5+e\alpha
\right)\right]\ , \label{gf19}
\ee
where $\phi\equiv e\partial_2\alpha|_{y=0}2y_0$ is the phase factor
implied by the physical distinction between $\psi_R$ and $\psi_L$, as fields
defined in edges separated by $2y_0$, a simple check shows us that
(\ref {gf17}) in fact satisfies all the above conditions.

We have not considered, in the microscopic derivation, the presence of two-body
interactions. Indeed, as far as the wavefunctions at the edges
are delocalized in the $x$ direction, their low electric charge densities
make the Coulomb repulsion generate logarithmic effects on the spectrum
\cite{giovanazzi}.
The Coulomb potential may be relevant,
however, in situations where we have localized states, induced by specific
configurations of the tunneling amplitude.

\section{Charge Trapping and Conductivity at the Interface}
Let us explore some of the consequences of the effective model of tunneling,
given by
(\ref {gf17}). We will consider first the case $a_0=a_1=a_2=0$. This model
may exhibit the phenomenon of charge fractionalization \cite{jackiw-rebbi},
intrinsically
related to the global behaviour of the external fields $f_1(x)$ and $f_2(x)$.
A gradient expansion of the fermion current, $\ran{J^\mu}$, computed assuming
$f_1$ and $f_2$ as slowly variating fields in space-time, yields \cite{gw}
(see appendix)
\begin{equation}
\ran{J^\mu(x)}={1\over (2\pi)}\ \epsilon^{\mu\nu}\ \epsilon^{ab} \ {f_a\
\p_\nu\ f_b\over |f|^2}={1\over (2\pi)}\ \epsilon^{\mu\nu} \ \p_\nu
\left[\tan^{-1} \left({f_2\over f_1}\right)\right] \ .\label{ct1}
\ee
For a configuration given by $f_1=\epsilon$, $\epsilon \rightarrow 0$ and
$f_2=A\tanh(\lambda x)$, we obtain, from the expression for $\ran{J^\mu}$,
a total charge $+1/2$ or $-1/2$, localized near $x=0$, when $\epsilon
\rightarrow 0^+$ or $\epsilon \rightarrow 0^-$, respectively. The role of
$f_1$ is limited to providing a regularization of (\ref {ct1}). The two
possibilities
for the total charge are associated with a zero mode in the spectrum, which by
its turn implies a doubly degenerate state: the occupied zero mode, with
charge $1/2$ and the unoccupied one, with charge $-1/2$. According to the
discussion of section III, this profile of $f$ corresponds to the
modulating potential
\begin{equation}
g(x)=2A\tanh(\lambda x)\sin(2 \bar k x) \ . \    \label{ct2}
\end{equation}
This potential could, in principle, be manufactured with the techniques used
in the fabrication of microstructures. A difficulty is that the spatial period
of (\ref {ct2}) occurs at not yet very manageable scales ($\sim 100 \AA$), by
present day technology. However, it is possible that in disordered
interfaces (which must contain, in Fourier space, modes close to $2 \bar k$),
modulated in adequate scales by a function qualitatively similar to $\tanh(
\lambda x )$, we have a more pratical way to observe the occurrence of fermion
number fractionalization.

Let us now investigate the introduction of perturbing gauge fields in the
system. There is a curious Aharonov-Bohm effect, which shows how a magnetic
flux may control the amount of charge localized at the interface. In order to
see
it, we consider $a_0=0$, $f=const.$ and a magnetic flux $\Phi$, confined in the
interior of a very thin imaginary solenoid crossing the plane at the point
$(x,y)=(0,0)$. In this way, the exponential factor which appears in
(\ref {gf17}) is $\exp(-ie\Phi \gamma^5)$, for $x=0^-$ and
$\exp(ie\Phi \gamma^5)$, for $x=0^+$. According to (\ref {ct1}), this
discontinuity will induce an accumulation of charge $\Phi / \pi$ at $x=0$.

It is important to remark that equation (\ref {ct1}), while giving a good
evaluation of the total amount of charge trapped by variations of
the tunneling amplitude, must be replaced by a sharper computation if we
want to know the size of the region where most of the charge is confined.
This may be found through exact solutions of the Dirac equation, in special
cases. In a ``soliton" profile, as $f=iA\tanh(\lambda x)$, the zero mode may
be explicitely computed \cite{bell} and the localization region inferred to be
\begin{equation}
\delta \sim \lim_{x \rightarrow \infty} \left( {2 x v} \over { \int^x_0
c|f(y)|dy} \right) \ . \ \label {ct3}
\end{equation}

We also note that in the situations we have been discussing, the charge
concetrated at the interface will have an interesting internal structure:
the relation between Fourier space and coordinate space (the $y$ direction)
will make any charge distribution have two equivalent and disjoint pieces,
localized at the edges R and L. The numerical analysis of section VI will
show this effect very clearly.

Since we know how the system is coupled to gauge fields, we can compute the
conductivity at the interface, as a response to small electric fields.

The current density which crosses the interface, $j_\bot$, flowing from one
edge
to the other, may be calculated through
\begin{equation}
j_\perp=\frac{i}{2y_0}\frac{\delta\ln Z[a_2]}{\delta a_2}=\frac{i}{2y_0} \left.
\frac{\delta\ln Z[a_2+\alpha]}{\delta\alpha}\right |_{\alpha=0}\ \ ,
\label{ct4}\ee
where $\alpha=\alpha(x,y)$ and
\begin{equation}
Z=\int D\bar\psi D\psi exp\left[ i\int d^2x {\cal L}_{eff} \right]
\label{ct5}
\ee
is the generating functional of model (\ref{gf17}).
We can now use the chiral anomaly \cite{dittrich} of model (\ref {gf17}) to
find
\begin{equation}
Z[a_2+\alpha]=Z[a_2]\cdot exp\left[i\int d^2x \frac{e^2\alpha y_0}{2\pi}
\epsilon_{\mu\nu}F^{\mu\nu}\right]~~,  \label{ct6}
\ee
where $F^{\mu\nu}=\partial^\mu a^\nu-\partial^\nu a^\mu$. In this way,
from (\ref{ct6}) and (\ref{ct4}), we obtain
\begin{equation}
j_\perp=-\frac{e^2}{2\pi}E_x \label{ct7}
\ee
Above, $E_x$ is the $x$ component of the electric field.
We may also study the influence of a small electric field, pointing in the $y$
direction. In this case, we have
\begin{equation}
a^0=a^1=0~~~\mbox{and}~~~a_2=E_y t\ , \label{ct8}
\ee
where $E_y$ represents the $y$ component of the electric field.
Applying (\ref{ct1}) to this problem, we get
\begin{equation}
\langle J^1\rangle_x=\frac{1}{2\pi} \epsilon^{10}\partial_0(-2ea_2y_0)
=\frac{e}{2\pi}E_y2y_0 \ . \ \label{ct9}
\ee
Therefore, the above relations tell us that the response to external electric
fields is given (in the usual units) by the following conductivity
tensor
\begin{equation}
\sigma=\left(
\begin{array}{lr}  0 & \sigma_{xy} \\
                -\sigma_{xy} & 0
\end{array} \right)
{}~~~\mbox{with}~~~\sigma_{xy}=\frac{e^2}{h}~~,
\label{ct10}
\ee
showing the quantization of the Hall conductivity at the interface.
It is interesting to note that this result is strongly related to the
imposition
of gauge invariance and is independent on the specific configuration of the
tunneling amplitude, even when it is associated to localized states. The above
argument may be regarded as an exact version, worked out for a particular
problem, of the fundamental works of Halperin \cite{halperin} and
Laughlin \cite{laughlin}, , since here we do not have to take any averages over
magnetic fluxes.

\section{Lattice Effects}
The quantum Hall effect is observed in approximately two-dimensional electron
systems, confined at the interface of semiconductor devices as Si-SiO$\sb 2$
or GaAs-Ga$\sb{1-x}$Al$\sb x$As.
The latter has been preferred in recent years due to its high-mobility
parameters, allowing for very precise measurements of conductivity and other
physical quantities \cite{qhe}.

Let us consider for a study of lattice effects, the two-dimensional electron
gas as having a certain thickness $2L_\bot$, experimentally found to be around
$50 \AA$, and interacting with a three-dimensional atomic lattice of
macroscopic
size, in fact much larger than any relevant characteristic length in the
process of tunneling. The lattice structure is relatively complex. The GaAs
lattice, for instance, has a zincblende structure, consisting of two
interpenetrating fcc sublattices, one of Ga and the other of As atoms. The
inter-atomic distance, $s$ , is close to $5 \AA$. We will study the effects of
the
lattice on the tunneling, through an approach analogous to that of Takayama et
al. \cite{takayama} for the case of polyacetylene.

The fortunate fact that the magnetic length is nearly $100 \AA$ in
the quantum Hall
effect, means that we can, in order to estimate couplings, simplify the
discussion assuming the lattice to have a simple cubic structure composed of
interlinked atoms
through an harmonic potential. Therefore, the lattice
potential may be expressed as
\begin{equation}
V(\vec x)=\sum_i \left[ V_0(\vec x - \vec x_i) - \xi_j(\vec
x_i)\partial_jV_0(\vec x - \vec x_i) \right] \ , \label {le1}
\ee
a sum of single atomic potentials centered at sites $\vec x_i=s(n_1\hat x +
n_2\hat y + n_3\hat z)$, with $n_1$, $n_2$, $n_3$ integers,
where we consider distortions as represented by
$\vec \xi(\vec x_i)$, in a linear approximation. We will be interested in
static distortions of the lattice, so that we can write the total Hamiltonian
of the system (quasi one-dimensional interface + lattice) as
\begin {equation}
H=H^I_{eff} + {1\over 4}K\sum_{<i,j>}\left( \vec \xi(\vec x_i) - \vec \xi(\vec
x
_j) \right)^2 + \int d^3\vec x \phi^+(\vec x)\phi(\vec x)V(\vec x) \ , \
\label {le2}
\ee
where $H^I_{eff}$ is given by (\ref{md43}), $K \sim  10^{-6} (\AA)^{-3}$
is the force constant associated to distortions, and $\phi(\vec x)$ is defined
from (\ref{md20}), according
to $\phi(x,y) \rightarrow \phi(x,y,z)=\left(2 /(  \pi L^2_\bot )
\right)^{1\over
4}exp\left(-\left(z /  L_\bot \right)^2  \right)\phi(x,y)$ ,
which corresponds to three-dimensional wavefunctions confined in $|z|<L_\bot$.
The sum in (\ref {le2}) is carried over nearest neighbors. We are also
assuming the absence of additional external gauge fields.

A convenient approximation is to replace sums over sites in (\ref{le1}) and
(\ref{le2}) by integrals. We find, then
\begin{equation}
V(\vec x)={1\over{s^3}} \int d^3\vec x V_0(\vec x) + {1\over{s^3}} \int d^3
\vec x_i  \partial_j \xi_j(\vec x_i)  V_0(\vec x - \vec x_i) \label {le3}
\ee
and
\begin{equation}
H=H^I_{eff} + {K\over{2s}}\int d^3 \vec x \partial_i \xi_j (\vec x) \partial_i
\xi_j (\vec x) + \int d^3 \vec x \phi^+(\vec x)\phi(\vec x)V(\vec x) \ . \
\label {le4}
\ee
As an estimate, we can write $V_0(\vec x)=-4 \pi Ze^2s^2 \delta^3(\vec x)$,
where
$e^2=1/137$ is the fine structure constant and $Z \simeq 2$ is an effective
atomic number.
Substituting the above
quantities in $V(\vec x)$, equation (\ref {le3}), we obtain
\begin{equation}
V(\vec x)=-{{4 \pi Ze^2}\over s}-{{4 \pi}\over s}Ze^2 \partial_j \xi_j(\vec x)
\ . \ \label {le5}
\ee
The first term on the RHS of (\ref {le5}) is a constant and may be shifted to
zero. As we noticed in section III, only certain Fourier components of $V(\vec
x)$ will be relevant in the tunneling. This amounts to considering a new set of
fields, $\vec \sigma (\vec x)$ and $\varphi(\vec x)$, given by
\begin{equation}
\vec \xi(\vec x) \simeq \vec \sigma(\vec x) \cos(2 \bar k x + \varphi(\vec x))
\ . \ \label {le6}
\ee
{}From (\ref {le6}) we get,
neglecting gradients of $\vec \sigma$ and $\varphi$,
\bea
\partial_j \xi_j(\vec x)=-2 \bar k \sigma_1(\vec x)\sin(2 \bar k x +
\varphi(\vec x)) +
 \cos(2 \bar k x +
 \varphi(\vec x))\partial_j\sigma_j(\vec x) + \nonumber \\
-\sigma_j(\vec x)\sin(2 \bar k x + \varphi(\vec x))\partial_j\varphi(\vec x)
 \simeq
-2 \bar k \sigma_1(\vec x)\sin(2 \bar k x + \varphi(\vec x)) \ . \
\label {le7}
\eea
Therefore, we see that only $\sigma_1$ must be considered in our analysis. On
the other side, the quadratic term in $\vec \xi$ in the Hamiltonian (\ref
{le4})
turns out to be (restricted to $\sigma_1$)
\bea
{K\over{2s}}&\int&d^3 \vec x \partial_j\left (\sigma_1\cos(2\bar k x +\varphi
)\right)\partial_j\left(\sigma_1\cos(2 \bar k x + \varphi)\right)= \nonumber\\
&=&{K\over{2s}}\int d^3 \vec x\left[(2\bar k)^2\sigma_1^2\sin^2(2 \bar k x
+\varphi)+
(\vec \nabla \sigma_1)^2\cos^2(2 \bar k x
+ \varphi)+\right. \nonumber \\
&+&\left.(\vec \nabla \varphi)^2\sin^2(2 \bar k x + \varphi)
- \partial_j\sigma_1
(2 \bar k\delta_{1j}+\partial_j \varphi)\cos(2 \bar k + \varphi)\sin(2 \bar k x
+\varphi)\right ] \nonumber \\
&\simeq& {K\over{4s}}\int d^3 \vec x\left [ (2 \bar k)^2 \sigma^2_1 +
(\vec \nabla \sigma_1)^2 + (\vec \nabla \varphi)^2\right ] \ , \ \label {le8}
\eea
where, assuming that $\sigma_1$ and $\varphi$ are dominated by slow modes, we
have substituted the above
trigonometric functions by their averaged values.

Defining the complex field $\sigma=\sigma_1\exp(i\varphi)$ and using the
approximations (\ref {le7}) and (\ref {le8}), we find, for the total
Hamiltonian of the system,
\bea
H&=&H^I_{eff}+{K\over{4s}}\int d^3 \vec x \left[(2 \bar k)^2|\sigma|^2 +
(\partial_j\sigma)^+(\partial_j\sigma) \right] + \nonumber \\
&+&\left[ {{4 \pi Ze^2 \bar k i}\over s}\int d^3 \vec x \phi^+(\vec x)\phi
(\vec x)\exp(2i \bar k x)\sigma(\vec x) +\hbox{H.c.} \right] \ . \ \label {le9}
\eea
It is useful to work in Fourier space, through
\begin{equation}
\sigma(\vec x)={1\over{(2\pi)^{3/2}}}\int d^3 \vec q \exp( i\vec q \cdot \vec
x)
\tilde \sigma(\vec q) \ . \ \label {le10}
\end{equation}
We now define a variational ansatz for the Hamiltonian (\ref {le9}): let us
consider a class of fields $\tilde \sigma(\vec q)$, parametrized by $\Delta_1,
\Delta_2 \geq 0$, according to
\begin{equation}
\tilde \sigma(\vec q)= \left\{
\begin{array}{ccccc}
\tilde\sigma(q_x,0,0)&\hbox{if}&  |q_y|\leq\Delta_1/\ell&\hbox{ ,
}&|q_z|\leq\Delta_2/L_\bot \\
0, &\hbox{otherwise.}&
\end{array}
\right.            \label {le11}
\end{equation}
This means , roughly, that we are taking the distortion field $\sigma(\vec x)$
as diferent from zero only in a certain neighborhood of the interface, given,
in Fourier space by the two variational parameters $\Delta_1$ and $\Delta_2$.
Substituting (\ref {le11}) in (\ref {le9}) we find, after straightforward
computations, the Hamiltonian
\bea
H&=&H^I_{eff}+{{K\Delta_1\Delta_2}\over{s \ell L_\bot}}\int dx
\left[(\partial_x
\sigma(x))^+(\partial_x\sigma(x))+\eta|\sigma(x)|^2 \right] + \nonumber \\
&+&\left[icf(\Delta_1,\Delta_2)\int dx \psi^+_L(x)\psi_R(x)\sigma(x) +
\hbox{H.c.} \right] \ , \   \label {le12}
\eea
where
\begin{equation}
\sigma(x)={1\over{(2 \pi)^{1/2}}}\int dq \exp(iqx) \tilde \sigma(q,0,0) \ , \
\label
{le13}
\end{equation}
\begin{equation}
\eta={\Delta_1^2\over{3 \ell^2}}+{\Delta_2^2\over{3 L_\bot^2}}+(2 \bar k)^2
\ , \ \label {le14}
\end{equation}
\begin{equation}
c={{2 \bar k Ze^2 (2 \pi)^{1/2}}\over s} \exp\left(-(\bar k \ell)^2
\right) \ , \  \label {le15}
\end{equation}
and
\begin{equation}
f(\Delta_1,\Delta_2)=\int^{\Delta_1/ \ell}_{- \Delta_1/ \ell} dq_1 \exp \left(
-{\ell^2 \over 4}q^2_1 \right)\int^{\Delta_2/L_\bot}_{-\Delta_2/L_\bot} dq_2
\exp \left( -{L_\bot^2 \over 8}q^2_2 \right)   \ . \   \label {le16}
\end{equation}
In order to study the variational problem, we will consider the path-integral
formalism, defining the generating functional
\begin{equation}
Z=\int D \bar \psi D \psi D \sigma^+ D \sigma \exp \left(iS[\psi,\sigma]
\right)
\ , \   \label {le17}
\end{equation}
where
\bea
S[\psi,\sigma]&=&\int d^2 x \bigg\{i \bar \psi(\gamma^0\partial_0 + v\gamma^1
\partial_1)\psi + \bar \psi \left(t_1+it_2\gamma^5-cf(\Delta_1,\Delta_2)
|\sigma|\exp(-i\varphi \gamma^5) \right)\psi+ \nonumber \\
&-& {{K\Delta_1\Delta_2}\over{s \ell L_\bot}}\left[(\partial_1 \sigma)^+
(\partial_1 \sigma)+ \eta |\sigma|^2 \right] \bigg\} \ . \ \label {le18}
\eea
We will simplify our discussion
, assuming that $t_1,t_2=0$, or in other words, that we have a very clean
interface, without any degree of disorder or modulating potentials.
We can integrate over the fermion fields in (\ref {le17}), neglecting
variations
of $|\sigma|$ and $\varphi$. This may be considered as the first term in a
gradient expansion of the fermion determinant. We get, then,
\begin{equation}
Z_{eff}=\int D \sigma^+ D\sigma \exp\left(iS_{eff}[\sigma]\right) \ , \
\label {le19}
\end{equation}
with
\begin{equation}
S_{eff}=-\int d^2 x \left\{ {{K\Delta_1\Delta_2}\over{s \ell \L_\bot}}
\left[(\partial_1\sigma)^+(\partial_1\sigma)+\eta|\sigma|^2\right]+V_{eff}(|\sigma|^2)
\right\} \ , \ \label {le20}
\end{equation}
where
\begin{equation}
V_{eff}(|\sigma|^2)={{(cf(\Delta_1,\Delta_2)|\sigma|)^2}\over{2 \pi v}}
\left[\ln\left({{2cf(\Delta_1,\Delta_2)|\sigma|}\over{v \bar k}}\right)^2+1
\right] \ . \ \label {le21}
\end{equation}
In the computation of (\ref {le21}) it is important to consider the presence
of a cutoff at $\bar k $ in the fermion theory. The variational strategy
is to find extremes of (\ref {le20}), in the space of configurations of
$\sigma(x)$ and also in the space of parameters $\Delta_1$ and $\Delta_2$.
Let us perform this analysis in two steps: first, we consider a fixed pair
$(\Delta_1,\Delta_2)$ in order to find the (constant) field $\bar \sigma
(\Delta_1,\Delta_2)$ which extremizes $S_{eff}$. Second, we look for
extremes of $S_{eff}\left[\bar \sigma \left(\Delta_1,\Delta_2\right) \right]$,
in the space $\left(\Delta_1,\Delta_2\right)$.

We obtain, in the first step, the gap in the fermion spectrum,
\begin{equation}
|cf\left(\Delta_1,\Delta_2\right)\bar \sigma|=v \bar k \exp\left[
-1-{{\pi v K \Delta_1 \Delta_2 \eta}\over { c^2 s \ell L_\bot}} \left({1\over
{f\left(\Delta_1,\Delta_2\right)}}\right)^2\right] \label {le22}
\end{equation}
and the effective action, evaluated for $\bar \sigma\left(\Delta_1,\Delta_2
\right)$,
\bea
&&S_{eff}\left[\bar \sigma \left(\Delta_1,\Delta_2\right)\right]=-
{{K\Delta_1\Delta_2\eta}\over{2 s \ell L_\bot}}|\bar \sigma|^2 \nonumber \\
&&=-{{E\Delta_1\Delta_2\left(F\Delta_1^2+G\Delta_2^2+H\right)}\over
{\left(f\left(\Delta_1,\Delta_2\right)\right)^2}}\exp\left[-2-{{I\Delta_1
\Delta_2\left(F\Delta_1^2+G\Delta_2^2+H\right)}\over
{\left(f\left(\Delta_1,\Delta_2\right)\right)^2}}\right] \ , \
\label {le23}
\eea
where
\begin{equation}
E={{Kv^2\bar k^2}\over{8 s \ell L_\bot c^2}} \ , \ F={1\over{3 \ell^2}}
\ , \ G={1\over{3 L_\bot^2}} \ , \ H=(2 \bar k)^2 \ , \
I={{2 \pi v K}\over{c^2 s \ell L_\bot}} \ . \ \label{le24}
\end{equation}
It is readily seen that the second step in the variational analysis,
$\partial S_{eff} /\partial\Delta_1=\partial S_{eff}/\partial \Delta_2=0$,
leads to
\begin{equation}
{\partial \over{\partial\Delta_{1,2}}}\left[{{\Delta_1\Delta_2\left(F\Delta_1^2
+G\Delta_2^2+H\right)}\over{\left(f\left(\Delta_1,\Delta_2\right)\right)^2}}
\right]=0  \ , \   \label {le25}
\end{equation}
or
\begin{equation}
{{I \Delta_1 \Delta_2 \left( F \Delta_1^2 + G \Delta_2^2 + H \right)} \over
{\left( f \left( \Delta_1, \Delta_2 \right) \right)^2}} = 1 \ . \
\label {le26}
\end{equation}
In order to study the above equations, we have to define $\bar k$ and $\alpha$
 ($v$ depends on $\alpha$; see equation (\ref {md22})). Let us, then, consider
$\bar k= 3/(2 \ell)$, corresponding to an interface
with the edges separated by $\sim 3 \ell$, and $\alpha/ \ell = 1.02$.
With these definitions, we get $v \sim 5 \times 10^{-5}$, $c \sim ((2
\pi )^{1/2}/6)
\times 10^{-4}(\AA)^{-2}$ and $I \sim 4 \times 10^{-5}$.
A quick inspection shows that both (\ref {le25}) and (\ref {le26}) have
solutions and that the minimum of $S_{eff}[\bar \sigma(\Delta_1,\Delta_2)]$ is
obtained from (\ref {le26}).
This equation has, in fact, many solutions, which determine a curve in the
$(\Delta_1, \Delta_2)$ plane. An estimate yields an isotropic solution
$\Delta_1
\sim \Delta_2 \sim 10$.
If we look at (\ref {le11}), we see that the degenerecence in the solutions of
(\ref {le26}) could mean a ``torsion" in the lattice displacements, if the
parameters $\Delta_1$ and $\Delta_2$ were non-trivially dependent on $q_x$. We
assume, however, that the isotropic solution represents a mean field for the
fluctuations of $\Delta_1$ and $\Delta_2$.
Anyway, the gap, as given by (\ref {le22}) does not depend on the degenerecence
of $\Delta_1$ and $\Delta_2$: substituting (\ref {le26}) into (\ref {le22}), we
find two possible ``vacua",
\begin{equation}
cf\left(\Delta_1,\Delta_2\right)\bar \sigma \sim ^+_-
10^{-7} (\AA)^{-1}  \ , \  \label {le27}
\end{equation}
physically measurable as a gap at the interface.

We can, in the same way, find non-trivial solutions of the Euler-Lagrange
equations for the complex field $\sigma$. We will have, in general, solitons
which interpolate the vacuum values of $<\sigma>$. The transition
region is given by the square root of the ratio between the kinectic and mass
term coefficients in the action (\ref {le20}). This quantity is $\eta^{-1/2}$.
Using $\Delta_1$, $\Delta_2 \sim 10$, we have $\eta^
{-1/2} \sim 10 \AA$. Since $\sigma$ changes its sign in this specific
configuration,
we will have solitons carrying charges $^+_-1/2$ (compared to the
electron charge) propagating along the interface. Using equation (\ref {ct3})
we find that the charge will be spread, along the interface, in a region of
lenght $\sim 500 \AA$.
Here, as in the case of
polyacetylene, we can also have soliton-antisoliton (polarons) excitations.
One can suppose that the soliton states would be found as midgap excitations,
but since the gap is estimated to be small, it would be hard to observe
any kind of soliton production threshold through variations of the sample
temperature. We point, however, that there is a mechanism, related to
Coulomb repulsion, which raises the soliton energy up to more clearly
observable scales.
The argument is as follows.
As we have already noticed, the edges are associated to different
wave-numbers of the fields $\psi_R$ and $\psi_L$. This means that all the
charge-density
profiles of these excitations will be symmetrically displaced at opposite sides
of the interface.
In this way, the Coulomb repulsion will raise the soliton energy by $\sim
(e/4)^2/(2 \ell) \sim 3 \times 10^{-6} ( \AA )^{-1}$, which is
close to the gap between Landau levels ($\sim 4 \times 10^{-6} ( \AA
)^{-1}$). We see, therefore, that the Coulomb interaction, which was
playing a minor role in the theory so far,
has an important participation in the soliton
spectrum.

Another interesting point is the possibility of an enhanced electric
conductivity at the interface, via soliton excitations. The same phenomenon
was conjectured to be present in the polyacetylene, but in view of the
relatively small polymer filaments, the solitons probably do not contribute
directly to the conductivity \cite{roth,tsukamoto}. In our case, however,
the interface's length may be constructed many orders of magnitude larger
than the magnetic length, in such way that we could hope the solitons to
be relevant in the conductivity process.

\section{Numerical Analisys}
In order to test the accuracy  of the approximations made in sec.\ III we
performed a numerical investigation of the model of tunneling defined by
relation (3.1).

Let us consider the matrix elements of (3.1) in the basis of functions
$\varphi_n(x,y)$
, as given by (3.11), and put it in the folowing form
\begin{equation}
\label{matrix}
<n|H|m>= <n|H_0|m>+<n|H_I|m> \ , \
\end{equation}
where
\begin{eqnarray}
H_0&=& \frac{1}{2m}[(P_x+eBy)^2+P_y^2]+eV_1(y) \\
H_I&=& eV_2(x,y) \ . \
\end{eqnarray}

The elementary calculation of $<n|H_0|m>$ leads to a diagonal matrix
with its elements given by (3.10), with $p=0$.
In order to evaluate $<n|H_I|m>$ we must choose a specific modulating potential
barrier. Since we want to explore the occurrence of fermion number
fractionalization, we may take a function $f(x)$ with the same asymptotic
behaviour as the modulating potential considered in sec IV or that of the
predicted solitons of sec V. Some numerical improvement is obtained from the
consideration of
\begin{equation}
\label{modfun}
f(x)=-\left[\epsilon -ig~sgn(x)\left( 1-e^{-\lambda|x|}\right)\right] \ . \
\end{equation}
In (\ref {modfun}) we have considered $\lambda=(5\ell)^{-1}$
and $g=4/m\ell^2$.
We have performed the numerical diagonalization of (\ref {matrix}) considering
$L/\ell\simeq 900$, $\bar k = 2/ \ell$ and a ``smooth" interface, determined by
$(\alpha / \ell - 1) \sim 10^{-4}$. The eigenvalue
distribution is shown in fig.1. In this figure we show the gap due to the
presence of $V_2(x,y)$. The order of magnitude of the gap agrees with the
value obtained from (\ref {md43}), $\Delta \simeq cg$. The two points inside
the gap
correspond to the ``zero mode'' and to a spurious wavefunction (associated to
the periodic boundary conditions).

In order to obtain the charge distribution along  the interface we must
evaluate
\begin{equation}
\label{density}
|\psi_n(x,y)|^2= \sum_{ij} (c_i^{~n})^* c_j^{~n} \varphi_i(x,y)^*
\varphi_j(x,y) \ , \
\end{equation}
where $c_i^{~n}$ is the $i$ component of the $n^{th}$ eigenvector of
(\ref {matrix}). The result for the localized state (zero mode)
is shown in fig. 2a. On the other hand, a typical
delocalized state in the band, far away from the gap, is depicted In Fig. 2b.
Note that in the localized state we have peaks at both sides of the
interface. The integral of the two peaks is close to $1/2$,
so that at each side of
the interface there is an accumulated charge of $1/4$, measured in units of
the electron charge.
In the numerical computation the total charge is not exactly $1/2$, in view
of finite size effects, as was numerically verified.

It is interesting to analyse the ``vacuum'' structure of the
theory.
For this aim it is enough to
integrate expression (\ref {density}) in the $y$ direction.
We define, then, a projected charge density,
$|\psi_n(x)|^2 \equiv  \int_{-\infty}^{\infty}dy~|\psi_n(x,y)|^2$.
In the field theory context, the vacuum charge density is evaluated by means of
\begin{equation}
\rho=\langle  0|:\psi^{\dagger}(x)\psi(x):|0\rangle=
\frac{1}{2}\langle 0|[\psi^{\dagger}(x),\psi(x)]|0\rangle \ . \  \label
{vacuum}
\end{equation}
Note that in (\ref {vacuum}) not only the lower band but also the upper band is
considered in the calculation.
The numerical result, shown in fig 3, agrees with the field theory expression.
In this computation one observes that the states in the upper band cancel the
states in the lower band, in such a way that the charge density of the
``vacuum" turns out
to be determined only by the zero mode.
This fact suggests that the approximation of linearizing the energy near the
Fermi energy is indeed a very good aproximation.
This result also confirms that our matrix Hamiltonian is adequate
to obtain the physics at the interface, without interferences from the
bulk. Therefore, the equivalence of the two calculations presented above is a
clear
evidence that the ``Dirac sea'' of the field theory model reproduces accurately
the completely filled lower band.

\section{Conclusion}
We studied the tunneling across a quasi one-dimensional interface in the
integer
quantum Hall effect. The particular form of the electron spectrum at the
edges allowed a mapping from the microscopic definition of the interface to
a relativistic $(1+1)$-dimensional quantum field theory. Gauge fields and
lattice effects were considered in the description. Regarding the coupling
to gauge fields, the Hall conductivity was found to be quantized, independently
of the possible induction of localized states by a non-uniform tunneling
amplitude. We also obtained a peculiar Aharonov-Bohm effect, which shows the
influence of magnetic fluxes on the charge concetrated at the interface. The
study of interactions between edge excitations and a three-dimensional
lattice showed us a natural mechanism for the generation of fractionally
charged solitons propagating along the interface. They are associated
to topologically stable solutions of the Euler-Lagrange equations for a complex
scalar field. We point that these excitations may contribute strongly to the
$\sigma_{xx}$ component of the conductivity tensor. A numerical test supported
the field theory approximations in the case of charge trapping in a modulated
barrier.

We believe that an experimental investigation of the above predicted phenomena
is crucial for a further development of the theory, in the sense of a more
accurate quantitative description. Anyway, there are a certain number of
extensions of the present work which may motivate future
studies. These are related to different microscopic definitions of the
interface, a consideration of higher Landau levels, the introduction of spin
in the formalism (which may be relevant to the case of magnetic fields of
lower intensity) and a derivation of lattice effects, taking into account exact
crystal structures (a Monte Carlo numerical analysis with pseudopotentials
could provide some information on lattice distortions and solitons).

\acknowledgments
L. M. would like to thank F.D.M. Haldane, M. Moriconi and D. Tsui for
interesting conversations. This work was supported by CNPq (Brazil).

\appendix
\section{Fractional Fermion Number}
In order to make this paper self-contained, we will show in this
appendix how the fermion current (\ref {ct1}) may be obtained. We will
compute it through the adiabatic method, as outlined by
Goldstone and  Wilczek \cite{gw}.
A more rigorous approach may be found in ref. \cite{ns}.

Let us consider the following fermionic Lagrangian in $(1+1)$ dimensions
\begin{equation}
{\cal L}=\bar\psi i\partial\bs\psi+ c\bar\psi(f_1+i\gamma^5f_2) \psi \ ,
\label{a1}  
\ee
where $f_1$ and $f_2$ are classical external fields,
and the $\gamma$ matrices are defined in (\ref{em5}).
This model may exhibit the phenomenon of
fermion number fractionalization, according to the topology of the fields $f_1$
and
$f_2$.

The model is invariant under rotations in the  $(f_1,f_2)$ plane. This is
related to global chiral transformations. That is, considering
\begin{equation}
\bar\psi(f_1+i\gamma^5 f_2)\ \psi=|\vec f|\bar\psi\exp\left[ i\tan^{-1}
\left({f_2\over f_1}\right) \ \gamma^5\right]  \psi \ ,\label{a2}
\ee
where $|\vec f|=(f^2_1+f^2_2)^{1\over 2}$, we see that a rotation in the
plane of coordinates $(f_1,f_2)$ by an angle $\Theta$ gives
$\tan^{-1}(f_2/f_1)\to
\tan^{-1}(f_2/f_1)+\Theta$, which can be absorved by a global chiral
transformation
\bea
\psi&\to& \exp\left(-{i\over 2}\ \Theta \gamma^5\right)\ \psi \label{a3} \\
\bar\psi&\to& \bar\psi\exp\left(-{i\over 2}\ \Theta \gamma^5\right) \ . \
\label{a4}
\eea
The existence of this symmetry will be important to establish the topological
nature of   $\langle J^\mu\rangle$.

Let $x_0$ be a space-time point such that $f_1(x_0)\ne 0$ and
$f_2(x_0)=0$. If we consider $f_1$ and $f_2$ as slowly variating fields, we
can  calculate $\ran{J^\mu(x_0)}$ using a ``mass''
$-cf_1(x_0)$ and an interaction term  ${\cal L}_I=ic\bar\psi\gamma^5f_2\psi$,
as can be seen from (\ref{a1}). We may write
\begin{equation}
\ran{J^\mu(x_0)}=\ran{0|\bar\psi(x_0)\ \gamma^\mu\psi(x_0)|0}=
\lim_{\epsilon\to 0 \atop \epsilon^0< 0}\ran{0|T\bar\psi(x_0)\
\gamma^\mu\psi(x_0+\epsilon)|0} \ . \label{a6}
\ee
That is , defining  $x_1\equiv x_0$ and $x_2\equiv x_0+\epsilon$,
\begin{equation}
\ran{J^\mu(x_0)}=Tr[\Delta^\mu(x_1,x_2)] \ , \label{a7}
\ee
where $\Delta^\mu_{\alpha\beta}(x_1,x_2)=\gamma^\mu_{\alpha\sigma }\ran{0|T\bar
\psi_\sigma (x_1)\ \psi_\beta(x_2)| 0}$ and we are suppressing the limit
simbols
to simplify the notation.

Expanding (\ref{a7}) in a perturbative series, we find, up to the first order
in
$c$
\begin{equation}
Tr\left[\Delta^\mu(x_1,x_2)=\Delta^\mu_0(x_1,x_2)+c\int d^2x\ \Delta^\mu_0
(x_1,x)\ f_2(x)\
\gamma^5\gamma_\mu\Delta^\mu_0(x,x_2)\right] \ , \  \label{a8}
\ee
where
$\Delta^\mu_0(x_1,x_2)=\gamma^\mu S_0(x_1,x_2)=\gamma^\mu\ran{0|T\bar
\psi(x_1)\
\psi(x_2)|0} |_{f_2=0}$.
Now, since
\bea
\Delta^\mu_0(x_1,x)&=&\gamma^\mu(2\pi)^{-2}\int d^2 q_1\exp
[iq_1(x_1-x)] \ S_0(q_1) \label{a9} \\
f_2(x)&=&\int d^2k\exp(ikx)\ \tilde f_2(k) \ ,\label{a10}
\eea
where  $S_0(q)=(\not q-m)^{-1}$, we obtain
\bea
\ran{J^\mu(x_0)}&=&Tr\left[{c\gamma^\mu\over (2\pi)^4} \int
d^2q_1 d^2k d^2 q_2 d^2 x \ S_0(q_1)\ \tilde f_2(k)\ \gamma^5S_0(q_2)\
\exp[iq_1(x_1-x)]\right. \times\nonumber \\
&\times& \exp(ikx)\exp[iq_2(x-x_2)]\bigg] \ . \label{a11}
\eea
Integrating in $x$ and $q_1$ and substituting $q_2$ by $q$, we find
\bea
\ran{J^\mu(x_0)}&=&Tr\left[{c\gamma^\mu\over (2\pi)^2} \int d^2q d^2k
 \ S_0(k+q)\ \gamma^5 \tilde f_2(k)\ S_0(q)\ \exp[iq(x_1-x_2)]
\times\right.\nonumber \\
&\times& \exp (ikx_1)\bigg] \ . \label{a12}
\eea
In the limit  $x_1\to x_2=x_0$, we have
\begin{equation}
\ran{J^\mu(x_0)}=Tr\left[{c\gamma^\mu\over (2\pi)^2} \int d^2q d^2k
\ S_0(k+q)\ \gamma^5 \tilde f_2(k)\ S_0(q)\  \exp(ikx_0)\right] \ .\label{a13}
\ee
Substituting the expansion  $S_0(k+q)=S_0(q)+k^\nu {\p\over \p
q^\nu}  S_0(q)$ in the above expression, we get
\bea
\ran{J^\mu(x_0)}&=&Tr\left[{c\gamma^\mu\over (2\pi)^2} \int d^2q
 \ S_0(q)\ \gamma^5\  S_0(q)\ \underbrace{f_2(x_0)}_{=0} \right]+\nonumber\\
&+&Tr\left[-{ic\gamma^\mu\over (2\pi)^2}\ \p^\nu f_2(x_0)\int d^2q\ {\p\over
\p q^\nu}\ S_0(q)\ \gamma^5 S_0(q)\right]=\nonumber \\
&=&{ic\p^\nu f_2(x_0)\over (2\pi)^2} Tr\left\{\gamma^\mu \int d^2q
\left[{\gamma_\nu \over (q^2-m^2)}-{(\not q+m)(2q_\nu)\over (q^2-m^2)}
\right] \ \gamma^5{(\not q+m)\over (q^2-m^2)}\right\}=\nonumber \\
&=&{ic\p_\nu f_2(x_0)\over (2\pi)^2} Tr[\gamma^\mu\gamma^\nu\gamma^5] \int
d^2q\ {m \over (q^2-m^2)^2} \ . \label{a14}
\eea
Since $Tr[\gamma^\mu\gamma^\nu\gamma^5]=2\epsilon^{\mu\nu}$, we obtain
\begin{equation}
\ran{J^\mu(x_0)}={2i\epsilon^{\mu\nu}\ c\ \p_\nu\ f_2(x_0)\over (2\pi)^2}
\int d^2q\ {m\over (q^2-m^2)^2} \ . \label{a15}
\ee
Performing  a Wick rotation , $q_0\to iq_0$, the above integral yields
\bea
i\int d^2q \ {m\over (q^2+m^2)^2}&=&2\pi i\int^\infty_0 dq\
{mq\over (q^2+m^2)^2}=\nonumber \\
&=&-\pi im \int^\infty_0 dq\ {d\over dq} \left({1\over q^2+m^2}\right)
={\pi_i\over m} \ . \label{a16}
\eea
In this way, using $m=-cf_1(x_0)$, we find
\begin{equation}
\ran{J^\mu(x_0)}={1\over (2\pi)} \ \epsilon^{\mu\nu}\ {\p_\nu f_2(x_0)\over
f_1(x_0)} \ . \label{a17}
\ee

As we mentioned, the theory has  chiral invariance.
This means that we must write the current $\ran{J^\mu(x_0)}$ in a
chiral invariant way. In other words, it must be invariant under  rotations in
the $(f_1,f_2)$ plane.
Therefore, we are led to
\begin{equation}
\ran{J^\mu(x)}={1\over (2\pi)}\ \epsilon^{\mu\nu}\ \epsilon^{ab} \ {f_a\
\p_\nu\ f_b\over f^2}={1\over (2\pi)}\ \epsilon^{\mu\nu} \ \p_\nu
\left[\tan^{-1} \left({f_2\over f_1}\right)\right] \ .\label{a18}
\ee

It is interesting to note that from a perturbative calculation and taking into
account the chiral symmetry of the model, it was possible to find the
non-perturbative expression (\ref{a18}).

The dependence of $\ran{J^\mu}$ with the topology of the fields
$f_1$ and  $f_2$ may be clearly seen by considering, for example, the total
charge
\bea
Q&=&\int^\infty_{-\infty} \ran{J^0(x)} dx={1\over 2\pi}
\int^\infty _{-\infty} \p_1 \tan^{-1}\left({f_2\over f_1}\right)\ dx
=\nonumber \\
&=&{1\over 2\pi} \ \Delta\tan^{-1}\left({f_2\over f_1}\right)\ .\label{a19}
\eea

\begin{figure}
\caption{}
The energy as a function of the quantum number $k$. The unit of energy is
$10^{-4} (\AA)^{-1}$ while $k$ is measured in units of $2 \bar k$. The dots
inside the gap represent localized states. They were depicted there only
to help in the visualization of their energies.

\caption{}
The charge density at the interface, $| \psi (x,y) |^2$, and their level
curves. Fig. 2a) shows the profile of the charge trapped in the
potential barrier. Fig. 2b) shows the charge density of
a delocalized state.

\caption{}
The projected charge density at the interface, obtained from the integration
of the charge density in the $y$ direction. The total charge is close to
$-1/2$ (empty zero mode). The x coordinate is measured in units of the magnetic
length and all the wavefunctions considered in this computation were normalized
to $1$ in the interval $- 200 \leq x \leq 200$.

\end{figure}

\end{document}